\documentclass[english,prl,twocolumn,superscriptaddress,letterpaper]{revtex4-1}
\usepackage[english]{babel}
\usepackage[utf8]{inputenc}
\usepackage{amsmath}
\usepackage{amssymb}
\usepackage{graphicx}

\def\R{\mathbb R}
\def\C{\mathbb C}
\def\CP{\mathbb{CP}}
\def\equiv{:=}
\def\vec{\mathbf}
\def\d{\partial}
\def\mc{\mathcal}
\def\bc{\begin{center}}
\def\ec{\end{center}}
\def\nn{\nonumber}
\def\spaa#1#2{\langle#1\,#2\rangle}
\def\spbb#1#2{[#1\,#2]}
\def\VV#1#2#3{{\left\langle#1|#2|#3\right]}}
\def\Res{\operatorname{Res}}
\def\cut{\operatorname{cut}}
\def\tree{\operatorname{tree}}
\def\Lra{\,\Longrightarrow\,}

\begin{document}
\date{\today}
\author{Mads S{\o}gaard}
\affiliation{
Niels Bohr International Academy and Discovery Center, \\
Niels Bohr Institute, Blegdamsvej 17, DK-2100 Copenhagen, Denmark
}
\author{Yang Zhang}
\affiliation{
Institute for Theoretical Physics, ETH Z{\"u}rich \\
Wolfgang-Pauli-Stra{\ss}e 27, CH-8093 Z{\"u}rich, Switzerland \\
}

\title{Elliptic Functions and Maximal Unitarity}

\begin{abstract}
Scattering amplitudes at loop level can be reduced to a basis of linearly
independent Feynman integrals. The integral coefficients are extracted from
generalized unitarity cuts which define algebraic varieties. The topology of an
algebraic variety characterizes the difficulty of applying maximal cuts.  In
this work, we analyze a novel class of integrals whose maximal cuts give rise to
an algebraic variety with irrational irreducible components. As a
phenomenologically relevant example we examine the two-loop planar double-box
contribution with internal massive lines. We derive unique projectors for all
four master integrals in terms of multivariate residues along with Weierstrass'
elliptic functions. We also show how to generate the leading-topology part of
otherwise infeasible integration-by-parts identities analytically from exact
meromorphic differential forms.
\end{abstract}

\maketitle

Modern perturbative scattering amplitudes in gauge theories such as QCD are
calculated from general principles of analyticity and unitarity without
inspecting Feynman diagrams. By virtue of analyticity, amplitudes are
reconstructed from their singularity structure, while unitarity ensures that
residues factorize onto simpler objects. Starting from complex internal momenta
and three-point amplitudes whose form is entirely fixed by field theory
arguments, all trees are generated recursively
\cite{Britto:2004ap,Britto:2005fq} and then recycled for loops via unitarity
cuts \cite{Bern:1994zx,Bern:1994cg,Bern:1997sc,Bern:2000dn}.

At the one-loop level, all amplitude contributions can be extracted directly
from a small set of generalized unitarity cuts
\cite{Britto:2004nc,Forde:2007mi,Badger:2008cm}. The computation is fully
automated and has led to numerous precise predictions for collider physics. In
the past few years, steps toward an analogous framework at two loops known as
maximal unitarity have been reported; see
refs.~\cite{Kosower:2011ty,CaronHuot:2012ab} and subsequent generalizations
\cite{Johansson:2012zv,Johansson:2013sda,Sogaard:2013yga,Sogaard:2013fpa,
Sogaard:2014ila,Sogaard:2014oka}. Parallel developments at the level of the
integrand can be found in e.g.
refs.~\cite{Badger:2012dp,Zhang:2012ce,Badger:2012dv,Badger:2013gxa}.

The increase of complexity at two loops requires a sophisticated approach.
Previous works
\cite{Kosower:2011ty,CaronHuot:2012ab,Johansson:2012zv,Johansson:2013sda,
Sogaard:2013yga,Sogaard:2013fpa,Sogaard:2014ila,Sogaard:2014oka,
Huang:2013kh,Hauenstein:2014mda,Zhang:2014xwa}
lend credence to the belief of surmounting the problem by understanding the
underlying algebraic and differential geometry of scattering amplitudes. The
topology of the algebraic varieties associated with the maximal cuts examined so
far has been that of degenerate elliptic and hyperelliptic curves of which the
irreducible components are rationally parametrized Riemann spheres. The only
algebraic curve at one loop is a conic section from the triangle diagram. In
advanced problems such as maximal cuts in $D=4-2\epsilon$ dimensions and massive
internal particles, the irreducible components have nonzero genus, and it has
been an open problem for years to deal with this class of integrals. In this
paper, we present an analytic solution for genus-1 maximal cuts, based on
Weierstrass' elliptic functions. Our method is used to predict new partial
results for two-loop scattering with massive propagators. 

The first step of multiloop amplitude calculations is to employ integrand-level
reductions and integration-by-parts (IBP) relations to obtain a minimal basis of
Feynman integrals $\{I_k\}$. The amplitude can thus be written 
\begin{align}
\label{EQ_MASTER}
\mc A^{L\text{-loop}}_n = 
\sum_{k\in\text{Basis}}c_kI_k+\text{rational terms}\;,
\end{align}
and the $c_k$s are rational functions. The integrals are computed in dimensional
regularization once and for all.

The coefficients are extracted by applying generalized unitarity cuts
\cite{Britto:2004nc,Forde:2007mi,Badger:2008cm}. This operation is advantageous,
because a loop-level amplitude may be broken into trees,
\begin{align} 
\sum_{k\in\text{Basis}}c_k I_k\big|_{\cut} = 
\sum_{\text{states}}A_{(1)}^{\tree}
A_{(2)}^{\tree}\cdots A_{(m)}^{\tree}\;.
\end{align}

As factorization for general amplitude contributions is achievable only for
complex-valued momenta, the refined unitarity cut prescription involves contour
integrals,
\begin{align}
\int_{\R} dz\,\delta(z-q)\longrightarrow
\frac{1}{2\pi i}\oint_{C(q)}\!\frac{dz}{z-q}\;,
\end{align}
rather than delta functions. Here, $C(q)$ is a small circle centered at
$q\in\C$. In the multidimensional case, the integration contour $\Gamma$ is an
$n$-torus, and the integrand is a differential form, 
\begin{align}
\omega(z) = \frac{h(z)dz_1\wedge\dots\wedge dz_n}{f_1(z)\cdots f_n(z)}\;.
\end{align}
Let $\xi\in\C^n$ be an isolated zero of $f = (f_1,\dots,f_n)$. The multivariate
residue of $\omega$ at the pole $\xi$ is said to be nondegenerate if
$J(\xi) = \det_{i,j}\d f_i/\d z_j|_\xi\neq 0$. Explicitly,
\begin{align}
(2\pi i)^n\Res{}_{\{f_1,\dots,f_n\},\xi}(\omega) = 
\oint_{\Gamma}\omega(z) = h(\xi)/J(\xi)\;.
\label{EQ_NONDEGENERATE_RESIDUE}
\end{align} 
For the degenerate case, see e.g.
refs.~\cite{Sogaard:2013fpa,Sogaard:2014ila,Sogaard:2014oka}.

To ensure consistency of maximal cuts, it is necessary to take appropriate
linear combinations of residues to project out spurious terms which integrate to
zero on the real slice \cite{Kosower:2011ty}. The sources of spurious terms are
parity-odd Levi-Civita contractions and parity-even IBP reductions. Accordingly,
we demand that
\begin{align}
I_1 = I_2 \Lra I_1\big|_{\cut} = I_2\big|_{\cut}\;,
\end{align}
which imposes constraints on the weights. Resolving the constraints uniquely and
deriving the master integral coefficients is the essential task in maximal
unitarity.

The principal mathematical prerequisite for the remainder of this paper is the
theory of elliptic curves; see e.g. refs.~\cite{Silverman,Whittaker}. We will
study nondegenerate elliptic curves over the field of complex numbers, governed
by the Weierstrass equation,
\begin{align}
y^2 = 4x^3-g_2x-g_3\;, \quad g_2^3-27g_3^2 \neq 0\;,
\label{EQ_ELLIPTIC_CURVE}
\end{align}
where $g_2,g_3$ are called the Weierstrass invariants. The elliptic curve
\eqref{EQ_ELLIPTIC_CURVE} is topologically equivalent to a torus in $\CP^1$
that is naturally parametrized by Weierstrass' $\wp$-function and its first
derivative. Indeed, 
\begin{align}
\wp'(z;g_2,g_3)^2 = 4\wp(z;g_2,g_3)^3-g_2\wp(z;g_2,g_3)-g_3\;, 
\end{align}
is precisely of the form \eqref{EQ_ELLIPTIC_CURVE}. The Weierstrass
$\wp$-function is fixed once either $g_2,g_3$ or the half-periods
$\omega_1,\omega_2$ are specified. For compactness we will just write $\wp(z)$.
An essential property of the Weierstrass $\wp$-function is the addition law,
\begin{align}
\wp(z)+\wp(w)+\wp(z+w) = \frac{1}{4}\bigg(
\frac{\wp'(z)-\wp'(w)}{\wp(z)-\wp(w)}\bigg)^2\;.
\label{EQ_ADDITION_LAW}
\end{align}
Below we will frequently encounter the function
\begin{align}
\varphi(z,w)\equiv\frac{1}{2}\frac{\wp'(z)-\wp'(w)}{\wp(z)-\wp(w)}\;.
\end{align}
It is expressible in terms of the Weierstrass $\zeta$-function,
\begin{align} 
\varphi(z,w) = \zeta(z+w)-\zeta(z)-\zeta(w)\;.
\label{EQ_F_ZETA}
\end{align}
Moreover, since $\zeta'(z) = -\wp(z)$,
\begin{align}
\frac{d}{dz}\varphi(z,w) = \wp(z)-\wp(z+w)\;.
\label{EQ_ZETA_DIFF}
\end{align}
Finally, we introduce the Weierstrass $\sigma$-function. It is defined through a
logarithmic derivative,
\begin{align}
\frac{d}{dz}\log\sigma(z) = \zeta(z)\;,
\label{EQ_SIGMA_DEF}
\end{align}
and obeys the periodicity relation,
\begin{align}
\sigma(z+2\omega_k) = -e^{2\eta_k(z+\omega_k)}\sigma(z)\;, \quad
\eta_k\equiv\zeta(\omega_k)\;.
\label{EQ_SIGMA_RELATION}
\end{align}

\begin{figure}[!h]
\bc
\includegraphics{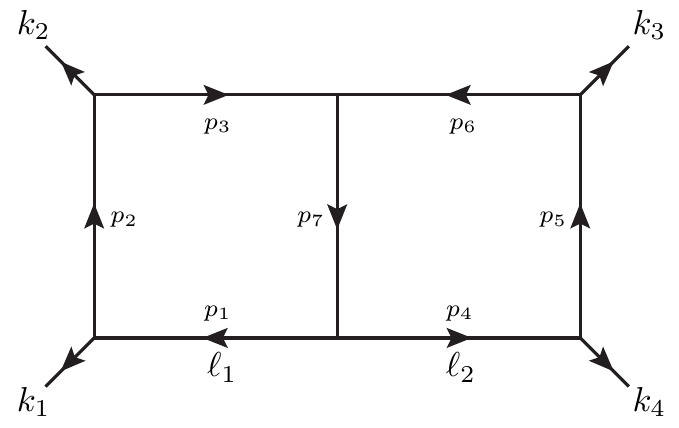}
\vspace*{-2mm}
\caption{\label{FIG_DBOX_INTERNAL_MASS}
The planar double box.}
\ec
\vspace*{-3mm}
\end{figure}

Our primary example is the planar double-box integral with internal masses
depicted in fig.~\ref{FIG_DBOX_INTERNAL_MASS}. Without loss of the main
features, we assume that external and internal lines are massless and massive, 
respectively. The outer-edge propagators carry mass $m_1$, while the particle in
the middle rung has mass $m_2$. The corresponding Feynman integral is denoted
$I$ and can easily be read off.

It is convenient to parametrize loop-momenta $\ell_1,\ell_2$ as
\begin{align}
\ell_1^\mu = {} & \alpha_1k_1^\mu+\alpha_2k_2^\mu+
\alpha_3\frac{s}{2}\frac{\VV{1}{\gamma^\mu}{2}}{\spaa{1}{4}\spbb{4}{2}}+
\alpha_4\frac{s}{2}\frac{\VV{2}{\gamma^\mu}{1}}{\spaa{2}{4}\spbb{4}{1}}\;, \nn \\
\ell_2^\mu = {} & \beta_1k_3^\mu+\beta_2k_4^\mu+
\beta_3\frac{s}{2}\frac{\VV{3}{\gamma^\mu}{4}}{\spaa{3}{1}\spbb{1}{4}}+
\beta_4\frac{s}{2}\frac{\VV{4}{\gamma^\mu}{3}}{\spaa{4}{1}\spbb{1}{3}}\;.
\end{align}
Simplifying the on-shell equations $p_1^2 = \cdots = p_6^2 = m_1^2$ yields 
$\alpha_1 = \beta_2 = 1$, $\alpha_2 = \beta_1 = 0$, 
$\alpha_3\alpha_4 = m_1^2t(s+t)/s^3$ and
$\beta_3\beta_4 = m_1^2t(s+t)/s^3$. The remaining cut equation is quadratic in
two variables, say, $\alpha_4$ and $\beta_4$. The solution is of the form 
$\beta_4 = \big(A(\alpha_4)+\sqrt{\Delta(\alpha_4)}\big)/B(\alpha_4)$.

The nondegenerate multivariate residue associated with the hepta-cut of $I$
easily follows from eq.~\eqref{EQ_NONDEGENERATE_RESIDUE},
\begin{align}
I|_{7-\cut}\propto\oint\frac{d\alpha_4}{\sqrt{\Delta}}\;,
\end{align}
and the constant of proportionality is not important for the argument. The
radicand $\Delta$ is a quartic polynomial,
\begin{align}
\Delta = q_0(\alpha_4-q)^4
+6q_2(\alpha_4-q)^2+4q_3(\alpha_4-q)+q_4\;,
\end{align}
for $q_i$s which are rational functions of kinematic invariants. The constant
$q = (m_2^2/(s-4m_1^2)-t/s)/2$ is designed to remove the cubic
term.

Generically, the four roots of $\Delta$ are distinct, so $\eta^2 = \Delta$
defines an elliptic curve. The structure of the roots is complicated, but it is
not necessary to solve for them explicitly. The elliptic curve is birationally
equivalent to the Weierstrass form \eqref{EQ_ELLIPTIC_CURVE}, with the
Weierstrass invariants,
\begin{align}
g_2 = (3q_2^2+q_0q_4)/q_0^2\;, \quad
g_3 = (q_0q_2q_4-q_0q_3^2-q_2^3)/q_0^3\;.
\end{align}
Via this birational transformation, the Weierstrass parametrization is found to
be of the form,
\begin{align}
\eta(z) = {} & \sqrt{q_0}(\wp(z)-\wp(z+u))\;, \quad
\alpha_4(z) = \varphi(z,u)+q\;,
\end{align}
where $u$ is the unique constant such that $\wp(u;g_2,g_3) = -q_2/q_0$ and
$\wp'(u;g_2,g_3) = q_3/q_0$. Invoking eq.~\eqref{EQ_ZETA_DIFF},
\begin{align}
\frac{d}{dz}\alpha_4(z) = \frac{1}{\sqrt{q_0}}\eta(z) \Lra
I[1]\big|_{7-\cut}\propto\oint dz\;.
\end{align}
Remarkably, all branch cuts are removed.

\begin{figure}[!h]
\bc
\includegraphics{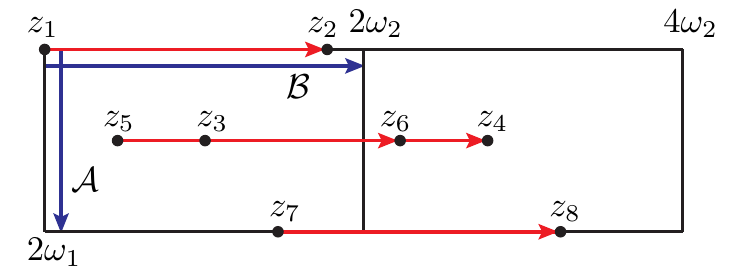}
\includegraphics{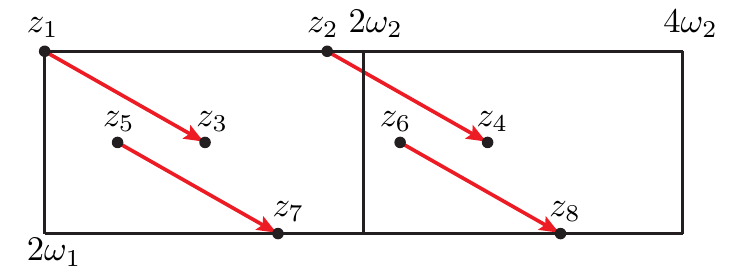}
\vspace*{-2mm}
\caption{\label{FIG_POLE_LOCUS}
The distribution of and relations among the eight poles
$(z_1,\dots,z_8)$ in the underlying lattice of the elliptic
functions. It can be seen that $z_{2i} = z_{2i-1}+z_2$ for $i = 1,2,3,4$ and 
$z_{i+2} = z_i+\omega_1+\omega_2$ for $i = 1,2,5,6$.}
\ec
\vspace*{-3mm}
\end{figure}

The half-periods of the torus associated with the elliptic curve are
$\omega_1,\omega_2$. For real $m_1,m_2,s_{12},s_{14}$ we choose $\omega_1$ to be
purely imaginary (with negative imaginary part) and $\omega_2$ to be real and
positive. The fundamental cycles $\mc A$ and $\mc B$ are depicted in
fig.~\ref{FIG_POLE_LOCUS}. We trivially find
\begin{align}
\oint_{\mc A}dz = 2\omega_1\;, \quad
\oint_{\mc B}dz = 2\omega_2\;,
\end{align}
and the scalar integrand has no poles. Evaluated on the hepta-cut, a generic
double-box numerator insertion is a polynomial in
$\alpha_3(z),\alpha_4(z),\beta_3(z),\beta_4(z)$. Let us examine the
$\alpha_4(z)$ insertion. The Weierstrass $\zeta$-functions in
eq.~\eqref{EQ_F_ZETA} can be integrated using
eqs.~\eqref{EQ_SIGMA_DEF} and \eqref{EQ_SIGMA_RELATION}, yielding
\begin{align}
\oint_{\mc A}dz\alpha_4(z) = {} &
2q\omega_1+\oint_{\mc A}dz\frac{1}{2}\frac{\wp'(z)-\wp'(u)}{\wp(z)-\wp(u)} 
\nn \\ = {} &
2q\omega_1+2(u\eta_1-\omega_1\zeta(u))\;,
\end{align}
and likewise for the $\mc B$ cycle. The poles of $\alpha_4(z)$ on the $z$-torus
are $z_1\equiv 0$ and $z_2\equiv -u$. By Laurent expansion,
\begin{align}
\oint_{\mc C_1}dz\alpha_4(z) = -2\pi i\;, \quad
\oint_{\mc C_2}dz\alpha_4(z) = +2\pi i\;,
\label{EQ_A4RESIDUES}
\end{align}
where $\mc C_i$ is a cycle around $z_i$. These residues sum to zero
by the Global Residue Theorem (GRT). The two poles of $\alpha_3(z)$, i.e.~the
two zeros of $\alpha_4(z)$, are located at $z_3\equiv z_1+\omega_1+\omega_2$ and
$z_4\equiv z_2+\omega_1+\omega_2$.~From the theory of elliptic functions, 
$\alpha_3(z) = \alpha_4(z-\omega_1-\omega_2)$, so $\alpha_3(z)$ is just a shift
of $\alpha_4(z)$. The shift leaves the fundamental cycle integrations invariant,
and the residues are also $\pm 2\pi i$. This analysis extends seamlessly to
linear insertions of $\beta_4(z)$ and $\beta_3(z)$. The two poles of
$\beta_4(z)$ are denoted by $z_5$ and $z_6$, and a short calculation reveals
that $z_6 = z_5+z_2$ and $\beta_4(z) = \alpha_4(z-z_5)$. An expression for $z_5$
can be found from the Weierstrass parametrization of $\beta_4(z)$. Similarly,
$\beta_3(z) = \alpha_4(z-z_7)$ and the poles are 
$z_7 = z_5+\omega_1+\omega_2$ and $z_8 = z_6+\omega_1+\omega_2$.
We denote the set of numerator poles as $\mc S=\{z_1,\dots,z_8\}$. In summary,
there are two fundamental cycles $\mc A,\mc B$ and eight residue cycles 
$\mc C_1,\dots,\mc C_8$ on the torus. Schematically,
\begin{align}
I_{0,0,0,0} \to {} & 
(2\omega_1,2\omega_2,0,0,0,0,0,0,0,0)\;, \\
I_{0,1,0,0} \to {} & 
(\mc A_{0,1,0,0},\mc B_{0,1,0,0},-2\pi i,2\pi i,0,0,0,0,0,0)\;, \nn \\
I_{1,0,0,0} \to {} & 
(\mc A_{1,0,0,0},\mc B_{1,0,0,0},0,0,-2\pi i,2\pi i,0,0,0,0)\;, \nn \\
I_{0,0,0,1} \to {} & 
(\mc A_{0,0,0,1},\mc B_{0,0,0,1},0,0,0,0,-2\pi i,2\pi i,0,0)\;, \nn \\
I_{0,0,1,0} \to {} & 
(\mc A_{0,0,1,0},\mc B_{0,0,1,0},0,0,0,0,0,0,-2\pi i,2\pi i)\;, \nn
\label{EQ_RANK_0_1}
\end{align}
with $I_{a,b,c,d}\equiv I[\alpha_3^a\alpha_4^b\beta_3^c\beta_4^d]$ and
\begin{alignat}{3}
\mc A_{1,0,0,0} = {} & \cdots = {} & \mc A_{0,0,0,1} = {} &
2q\omega_1+2(u\eta_1-\omega_1\zeta(u))\;, \nn \\
\mc B_{1,0,0,0} = {} & \cdots = {} & \mc B_{0,0,0,1} = {} &
2q\omega_2+2(u\eta_2-\omega_2\zeta(u))\;.
\end{alignat}
Note that $\eta_1\omega_2-\eta_2\omega_1 = i\pi/2$. The surprisingly simple
structure of the locus of poles is demonstrated in fig.~\ref{FIG_POLE_LOCUS}.

The weights associated with the fundamental cycles and the eight residues are
collected into a vector $\vec\Omega$,
\begin{align}
\vec\Omega = (\Omega_{\mc A},\Omega_{\mc B},\Omega_1,\dots,\Omega_8)^T\;.
\end{align}
We rewrite the remaining arbitrary one-dimensional integration contour in an
overcomplete basis of the first homology group of the $z$-torus with poles
excluded,
\begin{align}
&I[\Phi]\longrightarrow
\Omega_{\mc A}\oint_{\mc A}dz\Phi(z)+
\Omega_{\mc B}\oint_{\mc B}dz\Phi(z)
\nn \\[-1mm] &\hspace*{3cm}
+2\pi i\sum_{j=1}^8
\Omega_j\Res_{z=z_j}\Phi(z)\;,
\label{EQ_INTEGRAL_REDUCTION}
\end{align}
for a priori undetermined weights. The GRT implies that only seven of the
residues are independent.

The double-box topology with internal masses $m_1,m_2$ has four master
integrals, as can be verified from IBP identities generated by public computer
codes. The masters are typically chosen to be of the form 
$I_{m,n}\equiv I[(\ell_1\cdot k_4)^m(\ell_2\cdot k_1)^n]$. However, in practice
it proves advantageous to adopt master integrals with chiral numerator
insertions, for example,
\begin{align}
(I_1,\dots,I_4) = (I_{0,0,0,0},I_{0,1,0,0},I_{0,2,0,0},I_{0,1,1,0})\;.
\label{EQ_MASTERS}
\end{align}
Remarkably, there are five linearly independent constraints, leaving space for
precisely four master integral projectors. The constraints can be cast as a
matrix equation, $M\vec\Omega = 0$, for a coefficient matrix $M$ whose entries
are simply integers,
\begin{align}
M = \left(
\begin{array}{cccccccccc}
 1 & -1 & 0 & 0 & 0 & 0 & 0 & 2 & 0 & -2 \\
 0 & 0 & 1 & 0 & 0 & -1 & 0 & 1 & -1 & 0 \\
 0 & 0 & 0 & 1 & 0 & -1 & 0 & 1 & 0 & -1 \\
 0 & 0 & 0 & 0 & 1 & -1 & 0 & 0 & -1 & 1 \\
 0 & 0 & 0 & 0 & 0 & 0 & 1 & -1 & -1 & 1
\end{array}
\right)\;. 
\end{align} 
The origin of four of the constraints is conjugation symmetry, i.e. Levi-Civita
insertions which integrate to zero, whereas the last constraint reflects
left-right symmetry of the double-box diagram. All constraints from IBP
reduction are automatically satisfied. The rank-2 cut expressions in
eq.~\eqref{EQ_MASTERS} are of the form,
\begin{align}
I_{0,2,0,0} \to {} &
(\mc A_{0,2,0,0},\mc B_{0,2,0,0},-4\pi iq,4\pi iq,0,0,0,0,0,0)\;, \nn \\
I_{0,1,1,0} \to {} &
(\mc A_{0,1,1,0},\mc B_{0,1,1,0},r_+,r_-,0,0,0,0,-r_-,-r_+)\;,
\end{align}
where $r_\pm$ are simple functions of external invariants and 
$\mc A_{0,2,0,0}$ and $\mc B_{0,2,0,0}$ are easy to derive using
eq.~\eqref{EQ_ADDITION_LAW}. The analytic results for 
$\mc A_{0,1,1,0}$ and $\mc B_{0,1,1,0}$ are a bit more complicated.

Let $\mc M_i$ denote the projector which extracts $I_i$ and normalizes its cut
expression, respecting all constraints. The projectors can be written compactly
as solutions to inhomogeneous matrix equations. Therefore we construct the
$10\times 10$ matrix,
\begin{align}
F = \left(
\begin{array}{cccccc}
M & G & I_1 & I_2 & I_3 & I_4
\end{array}
\right)^T\big|_{\cut}
\end{align}
where transposition is with respect to the six blocks and 
$G = (0,0,0,0,0,0,0,0,0,1)$ implements the GRT. $F$ has full rank and all four
projectors are thus unique. Defining
\begin{align}
\left( 
\begin{array}{cccc}
\delta_1 & \delta_2 & \delta_3 & \delta_4
\end{array}
\right) = 
\left(
\begin{array}{c}
0_{6\times 4} \\
1_{4\times 4}
\end{array}
\right)\;,
\end{align}
we arrive at the result, $\mc M_i = F^{-1}\delta_i$. Our final formula for the
master integral coefficients is
\begin{align}
&c_i = 
\Omega_{\mc A}^{(i)}\oint_{\mc A}dz
\prod_{k=1}^6 A_{(k)}^{\tree}(z)+
\Omega_{\mc B}^{(i)}\oint_{\mc B}dz
\prod_{k=1}^6 A_{(k)}^{\tree}(z)
\nn \\[-1mm] &\hspace*{2cm}
+2\pi i\sum_{j=1}^8
\Omega_j^{(i)}\Res_{z=z_j}
\prod_{k=1}^6 A_{(k)}^{\tree}(z)\;.
\end{align}
Note that the products of tree-level amplitudes are implicitly summed over all
internal on-shell states in the theory.

An amazing property we developed is the relation between exact meromorphic
differential forms and IBP identities. Let $F$ be an elliptic function with
poles inside $\mathcal S$, so $dF=f dz$ is an exact 1-form on 
$\mathbf T^2 \setminus \mc S$. By Stokes' theorem, 
$\oint_{\mc A,\mc B}dF=\oint_{\mc C_i} dF = 0$. So from
eq.~\eqref{EQ_INTEGRAL_REDUCTION},
\begin{eqnarray}
I[f]=0+\cdots
\end{eqnarray}
is an IBP identity. Here, $\cdots$ stands for integrals with fewer than seven
propagators. Using the properties of Weierstrass' functions, 
\begin{align}
d(f_1(\alpha_4)\eta) = {} & \frac{f_1'(\alpha_4)
\Delta+\frac{1}{2}f_1(\alpha_4)\Delta'(\alpha_4)}{\sqrt{q_0}}dz\;, \\
d(f_2(\alpha_4)) = {} & \frac{f_2'(\alpha_4)(B(\alpha_4)
\beta_4-A(\alpha_4))}{\sqrt{q_0}}dz\;,
\end{align}
for arbitrary polynomials $f_1,f_2$. Hence, we get the IBPs,
\begin{align}
I[f_1'(\alpha_4)
\Delta+\tfrac{1}{2} f_1(\alpha_4)\Delta'(\alpha_4)] = {} & \cdots\;,
\label{EQ_IBP1} \\
I[f_2'(\alpha_4)(B(\alpha_4)\beta_4-A(\alpha_4))] = {} & \cdots\;.
\label{EQ_IBP2}
\end{align}
For example taking $f_2(\alpha_4)=\alpha_4$, the IBP
\begin{gather}
m_1^2t^2(s+t) I_{0,0,0,0}+
2s^4 I_{0,2,0,1}+
s^3t I_{0,2,0,0}+
2s^3t I_{0,1,0,1} \nn \\
+s^2t(2m_1^2-m_2^2+t) I_{0,1,0,0}+
2m_1^2st^2 I_{0,0,0,1} = \cdots 
\end{gather}
is obtained. It is verified that eqs.~\eqref{EQ_IBP1} and \eqref{EQ_IBP2} and
similar relations with respect to the flip symmetry generate {\it all} IBP
identities without doubled propagators for the massive double-box diagram. We
expect that this relation between meromorphic exact forms and IBP identities
would hold for other two-loop diagrams and lead to an extremely efficient
algorithm for generating IBPs analytically.

We remark that one is not obliged to work in the Weierstrass standard form.
Indeed, Weierstrass' elliptic functions are equivalent to the Jacobi elliptic
functions. The fundamental parameter of our torus, $\tau = \omega_2/\omega_1$,
is related to the elliptic modulus $k$ via the $j$-invariant.

The calculations presented here yield a highly nontrivial addition to the body
of evidence of the uniqueness conjecture of two-loop master integral projectors.
In particular, our work continues to suggest a very intimate connection between
the structure of maximal unitarity cuts and algebraic geometry and multivariate
complex analysis. This paper gives rise to a host of new exciting directions in
multiloop unitarity. The obvious extension is to formalize maximal cuts of
double-box integrals in $D$ dimensions. We expect this can be done by analytic
continuation. Our method presumably applies directly to the purely massless
double-box contribution to ten-gluon scattering \cite{CaronHuot:2012ab}. It
would be very interesting to understand the structure of maximal cuts which
define hyperelliptic curves, for example from the nonplanar double box, and more
generally, topologically nontrivial surfaces. We are also intrigued by
investigating the relation between maximal cuts and evaluation of master
integrals. These problems provide avenues for discovering further relations
between scattering amplitudes and areas of mathematics.

{\bf Acknowledgments:} We have benefited from discussions with N. Beisert,
N.E.J. Bjerrum-Bohr, S. Caron-Huot, P.H. Damgaard, H. Frellesvig, R. Huang, 
H. Ita, H. Johansson, D.A. Kosower and K.J. Larsen. MS and YZ are grateful to
the Institute for Theoretical Physics at ETH Z{\"u}rich and the HKUST Jockey
Club Institute, respectively, for hospitality during phases of this project. The
research leading to these results has received funding from the European
Research Council under the European Union's Seventh Framework Programme
(FP/2007-2013) / ERC Grant Agreement no. 615203. The work is partially supported
by the Swiss National Science Foundation through the NCCR SwissMAP.

\end{document}